\documentclass[superscriptaddress,reprint]{revtex4-2}
\usepackage{graphicx}
\usepackage{color}
\usepackage[dvipsnames]{xcolor}
\usepackage{bm}
\usepackage{amsmath}
\usepackage{amsfonts}
\usepackage{amssymb}
\usepackage{braket}
\usepackage[colorlinks,linkcolor=RoyalBlue, citecolor=RoyalBlue,
urlcolor=RoyalBlue]{hyperref}
\usepackage{dsfont}

\DeclareMathOperator{\Tr}{Tr}

\begin{document}

\title{Quantum-enhanced joint estimation of phase and phase diffusion}

\author{Jayanth Jayakumar}
\affiliation{Faculty of Physics, University of Warsaw, ul. Pasteura 5, 02-093 Warsaw, Poland}
\affiliation{Faculty of Mathematics, Informatics and Mechanics, University of Warsaw, ul. Banacha 2, 02-097 Warsaw, Poland}
\author{Monika E. Mycroft}
\affiliation{Faculty of Physics, University of Warsaw, ul. Pasteura 5, 02-093 Warsaw, Poland}
\author{Marco Barbieri}
\affiliation{Dipartimento di Scienze, Universit\`a degli Studi Roma Tre, Via della Vasca Navale 84, 00146 Rome, Italy}
\author{Magdalena Stobińska}
\affiliation{Faculty of Mathematics, Informatics and Mechanics, University of Warsaw, ul. Banacha 2, 02-097 Warsaw, Poland}

\begin{abstract} 
Accurate phase estimation in the presence of unknown phase diffusive noise is a crucial yet challenging task in noisy quantum metrology. This problem is particularly interesting due to the detrimental impact of the associated noise. Here, we investigate the joint estimation of phase and phase diffusion using generalized Holland-Burnett states, known for their experimental accessibility. These states provide performance close to the optimal state in single-parameter phase estimation, even in the presence of photon losses. We adopt a twofold approach by analyzing the joint information extraction through the double homodyne measurement and the joint information availability across all probe states. Through our analysis, we find that the highest sensitivities are obtained by using states created by directing all input photons into one port of a balanced beam splitter. Furthermore, we infer that good levels of sensitivity persist even in the presence of moderate photon losses, illustrating the remarkable resilience of our probe states under lossy conditions.
\end{abstract}

\maketitle

\section{Introduction}
\label{sec:introduction}

The advent of quantum metrology brings the promise of achieving better precision in measurements through the careful control of the probe quantum state~\cite{giovannetti2011advances,giovannetti2004quantum,pezze2018quantum}. Although its introduction was suggested under very idealised conditions~\cite{caves1981quantum,braunstein1992quantum,pezze2008mach,seshadreesan2011parity,yurke19862,bollinger1996optimal}, continuous progress has allowed to adopt it in real-life apparata, notably for the detection of gravitational waves~\cite{ligo2011gravitational,aasi2013enhanced}.
The foundation of quantum metrology rests on classical parameter estimation, where unknown parameters are inferred from a probability distribution describing the measurement~\cite{kay1993fundamentals,lehmann2006theory}. Particularly, the introduction of a quantum version, where parameters are encoded in the quantum state instead, has provided a proper framework for studying the measurement of physical quantities that do not directly correspond to quantum observables~\cite{helstrom1969quantum,holevo2011probabilistic}.

Further progress will continue towards the realisation of proper quantum sensors which will be characterized by their enhanced ability to detect physical quantities, as well as to monitor their changes. As for any reliable technology, such sensors should be able to operate outside protected environments by being made robust against perturbations and disturbing actions from the outside world. Since these detrimental effects may not be stationary, their assessment must occur concurrently with the actual sensing. However, it takes to introduce further parameters accounting for the nuisance~\cite{maccone2011beauty,banaszek2009quantum,dorner2009optimal,demkowicz2009quantum,cable2010parameter,knysh2011scaling,escher2011general}. Their presence is not just a matter of sheer inconvenience and reduction of the performance; the operation of the sensor, specifically of its probe state, requires reassessment~\cite{thekkadath2020quantum,oszmaniec2016random,huelga1997improvement,ouyang2021robust}.

Quantum  multiparameter  estimation is a powerful framework to understand such a problem and devise solutions, although it is unable to provide clear-cut answers unlike its single-parameter counterpart. It is based on finding a bound on the covariance matrix of the parameters, i.e., a multivariate quantum Cram\'er-Rao bound~\cite{albarelli2020perspective,szczykulska2016multi,demkowicz2020multi}, but there are no guarantees as to whether it could be achieved. In fact, the commutativity of the optimal measurements does dictate additional constraints on the saturability of the Cram\'er-Rao bound~\cite{albarelli2019evaluating,ragy2016compatibility}. These are specific to the investigated problem, but more general outlooks are often collected by inspecting examples.

For optical sensors, phase is the primary quantity which is considered. It could appear with fluctuations occurring on time scales shorter than, or of the same order as, the collection period. This is captured by a phase diffusion term whose amplitude constitutes a nuisance parameter~\cite{genoni2011optical,demkowicz2015quantum}. Due to the flexibility in encompassing disparate situations, phase with phase diffusion has attracted attention as an effective model in quantum sensing~\cite{szczykulska2017reaching,altorio2015weak,vidrighin2014joint,roccia2017entangling}. From a high-level perspective, the employ of probe states with larger dimensionality has been proven beneficial~\cite{vidrighin2014joint}; entangling measurements on multiple copies are a viable alternative~\cite{vidrighin2014joint,parniak2018beating, 
roccia2017entangling,hou2018deterministic,conlon2023approaching,wu2020minimizing,yuan2020direct}. However, most proposals do not seem to put forward practical considerations for the production of satisfactory states and for their detection. 

\begin{figure}[t]
\includegraphics[width=\columnwidth]{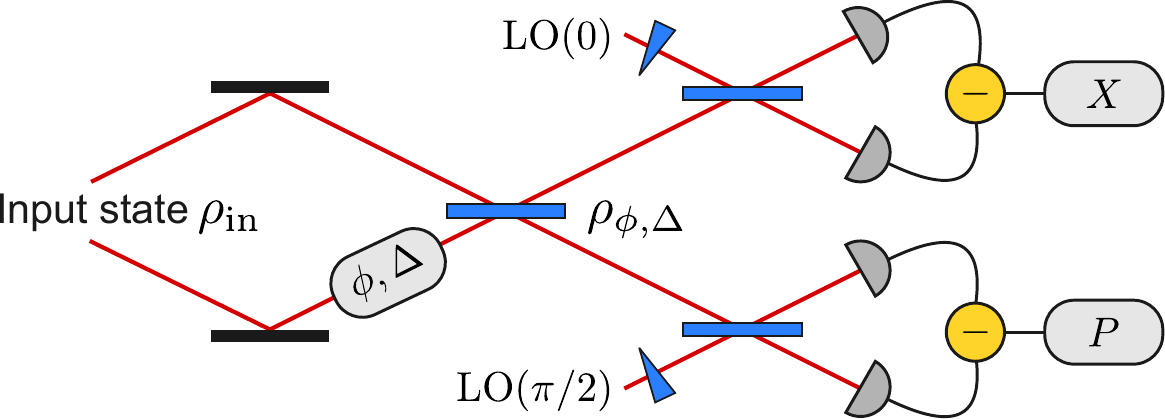}
\caption{The theoretical scheme for the joint estimation of phase $\phi$ and phase diffusion $\Delta$ involves a Mach-Zehnder interferometer followed by double homodyne detection. This setup provides the optimal strategy for the estimation of phase in the absence of phase diffusion with qubit probe states.}
\label{fig:setup}
\end{figure}

In this article, we make substantial progress on this front by presenting a thorough analysis of simultaneous phase and phase diffusion estimation based on homodyne measurements, which represent one of the most reliable techniques in optical detection, and grant high efficiency even without recurring to operation at cryogenic conditions~\cite{leonhardt1997measuring}. The probe states we consider are generalisations of those proposed by Holland and Burnett~\cite{holland1993interferometric}, obtained by interfering different Fock states on a beam splitter. This is an application of continuous variable detection to states that are almost always considered in tandem with photon counting schemes.

Our results demonstrate that a specific group of probe states, created by directing all input photons into one port of a beam splitter provides good performance. This applies to both efficiently extracting joint information from measurements and reducing absolute error on the parameters. The performance of these states are better than both the usual Holland-Burnett states and the N00N states. Remarkably, their advantages can actually persist in the presence of moderate loss. Furthermore, by exploiting the freedom to interfere different Fock states, we analyze the performance of a broad class of states. Interestingly, we find that among these states, the same class still yields the best performance.

\section{Results}

\subsection{The setting}
\label{sec:setting}

The basic features of a phase sensor can be captured by a Mach-Zehnder interferometer (MZI) in which the parameter to be estimated is the relative phase $\phi$ between its two arms. The detection scheme is a double homodyne detector, as shown in Fig.~\ref{fig:setup}.  As the phase shift may decohere while being measured, we need to introduce a second nuisance parameter that accounts for phase fluctuations. The overall goal is then to achieve the best possible precision on both parameters in a joint estimation. This setup was shown to provide the optimal strategy for phase estimation using probe states effectively described as equatorial qubit states~\cite{vidrighin2014joint}. These states also yielded reasonable precision for joint estimation. We will show that probes prepared in generalized Holland-Burnett (gHB) states, which are experimentally viable and live in higher-dimensional Hilbert spaces~\cite{mycroft2023proposal}, can provide significantly better precision.

The gHB states have been shown to perform close to the optimal estimation of phase in the presence of particle losses~\cite{thekkadath2020quantum}. They can be created by the action of a balanced beam splitter on a two-mode Fock state $\ket{n,N-n}$ defined as
\begin{align}
   |\Psi_{\mathrm{gHB}}(n,N-n)\rangle &{}= \mathcal{U}_{\mathrm{BS}}\ket{n,N-n} \nonumber\\
  &{} =\sum_{p=0}^N \mathcal{A}_N(n,p)\ket{p, N-p}
   \label{eqn:ghbstate}
\end{align} 
where $\mathcal{U}_{\mathrm{BS}}=\exp{[-i\frac{\pi}{4}(a'^{\dagger}b' + b'^{\dagger} a')]}$, $a'$ ($a'^{\dagger}$) as well as $b'$ ($b'^{\dagger}$) are the annihilation (creation) operators corresponding to the input modes, and $\mathcal{A}_{N}(n,p)=(-1)^n \sqrt{2^{-N} \binom{N}{n} \binom{N}{p}} \,
{}_2F_1\left(-n,-p;-N;2\right)$ are the Kravchuk coefficients~\cite{mycroft2023proposal}. As a special case, when we input equal numbers of photons into a balanced beam splitter, i.e., $n=N/2$, the Holland-Burnett (HB) states are created.

In the MZI, the input gHB state $\rho^{\mathrm{gHB}}=|\Psi_{\mathrm{gHB}}\rangle \langle\Psi_{\mathrm{gHB}}|$ acquires a phase $\phi$ through a unitary phase-shift operation $U_{\phi} =e^{-i \phi \, a^{\dagger}a}$, where $a$ ($a^{\dagger}$) is the annihilation (creation) operator corresponding to mode $a$. In the lossless conditions, the MZI output state is $\rho^{\mathrm{gHB}}_\phi=U_{\phi}\rho^{\mathrm{gHB}} U^{\dagger}_{\phi}$. We model random phase fluctuations with a phase diffusion channel $\Lambda_{\phi,\Delta}$ according to a Gaussian distribution $p_{\phi,\Delta}$, with mean $\phi$ and standard deviation $\Delta$. Now, the lossy MZI alters the input state $\rho^{\mathrm{gHB}}$ as follows:
\begin{equation}
\rho^{\mathrm{gHB}}_{\phi,\Delta}=\Lambda_{\phi,\Delta}(\rho^{\mathrm{gHB}} )=\int_{-\infty}^{\infty}d\phi' p_{\phi,\Delta}(\phi')U_{\phi'}\rho^{\mathrm{gHB}} U^{\dagger}_{\phi'}.
\label{eqn:krausrep}
\end{equation}
The phase diffusion channel erases the off-diagonal elements of $\rho^{\mathrm{gHB}}_{\phi,\Delta}$ leading to a significant loss of information about the phase, and in this way it hampers its estimation.

As a result of Eq.~(\ref{eqn:krausrep}), a gHB state is turned by the MZI into a mixed state
\begin{align}
   \rho^{\mathrm{gHB}}_{\phi,\Delta}(n,N-n)=\sum_{p,q=0}^{N}\mathcal{C}_{N}(n,p,q)\ket{p,N-p}\bra{q,N-q},
\label{eqn:outputghbstate}
\end{align}
where $\mathcal{C}_{N}(n,p,q)=\mathcal{A}_N(n,p)\mathcal{A}_N(n,q)e^{-i(p-q)\phi - \frac{\Delta^2}{2}(p-q)^2}$.

The performance of these states for the joint phase-phase diffusion estimation problem is captured by their Fisher information (FI). This figure of merit quantifies the amount of information about the parameters of a physical system that can be extracted asymptotically through a large number of repeated measurements~\cite{kay1993fundamentals,lehmann2006theory}. For a collection of $K$ parameters $\theta=\{\theta_i\}_{i=1}^K$, and the corresponding probability distribution of the measurement outcomes $p_n = p_n(\theta)$, the FI matrix $F_C$ can be expressed as $F_{C\, i,j}=\sum_n \tfrac{1}{p_n} \tfrac{\partial p_n}{\partial \theta_i} \tfrac{\partial p_n}{\partial \theta_j}$. The Cram\'er-Rao bound states that the maximal precision that can be attained in joint estimation of unbiased estimators $\tilde{\theta}$, expressed in terms of the covariance matrix $\gamma$, is set by the FI matrix and the number of experimental runs $M$, $\gamma \ge (M F_C)^{-1}$, where $\gamma_{i,j}=\langle (\tilde{\theta}_i -\theta_i)(\tilde{\theta}_j-\theta_j) \rangle$.

This classical precision limit can be beaten using quantum resources, up to the quantum Cram\'er-Rao (QCR) bound~\cite{helstrom1969quantum,holevo2011probabilistic}. A probability distribution of measurement outcomes performed on quantum states $\rho_{\theta}$ enables to determine the quantum Fisher information (QFI) matrix $F_Q$ that sets the ultimate QCR limit for the covariance matrix
\begin{align}
    \gamma \ge (M F_Q)^{-1}.
\end{align}
The QFI matrix can be defined in terms of the symmetric logarithmic derivative (SLD) operator $\mathcal{L}_j$ for parameter $\theta_j$, $\partial_{\theta_j} \rho_{\theta} = \tfrac{1}{2}(\mathcal{L}_j \rho_{\theta} + \rho_{\theta} \, \mathcal{L}_j)$. Then, $F_{Q \, i,j} = \frac{1}{2}\mathrm{Tr}(\rho_{\theta} \{\mathcal{L}_i, \mathcal{L}_j\})$, and $F_Q \ge F_C$.

In case of a single parameter quantum estimation, reaching the QCR bound is always possible by adopting the measurements given by the eigenvectors of the SLD. However, for a multiparameter problem, it is possible that the SLD operators corresponding to different parameters do not commute and the FI values for these parameters are maximized by incompatible measurements~\cite{heinosaari2016invitation,zhu2015information,ragy2016compatibility,demkowicz2020multi}. Then, a stronger condition for $\gamma$ applies, the Holevo Cram\'er-Rao bound \cite{hayashi2005asymptotic,holevo2011probabilistic}, and collective measurements may in general be required to saturate it. Nevertheless, the latter can coincide with the QCR bound for certain classes of problems, for example, if the optimal measurements satisfy the commutation condition \cite{matsumoto2002new,vaneph2013quantum,crowley2014tradeoff,suzuki2016explicit} (see Section~\ref{sec:methodsA} for details)
\begin{align}
    \mathrm{Tr}(\rho_{\theta} [\mathcal{L}_i, \mathcal{L}_j])=0.
    \label{eqn:comm_rel}
\end{align}

\subsection{Joint estimation of phase and phase diffusion}\label{sec:jointestimation}

We now study the problem of enhancing precision in multiparameter estimation of phase and phase diffusion towards the QCR bound. In our analysis, we adopt a twofold approach by introducing a couple of figures of merit. Our first figure of merit accesses the joint information extraction about the parameters through the measurement, quantified by
$\Upsilon=\Tr(F_C F^{-1}_Q)=F_{C\, 1,1}/F_{Q\, 1,1} + F_{C\, 2,2}/F_{Q\, 2,2}$. It is worth noting that this equality holds only if the off-diagonal elements $F_{C\: (Q) \: i,j}=0$, $i\neq j$, which implies that the estimators for the parameters are uncorrelated. The second figure of merit is given by the lower bound on the absolute error on the parameters or the sum of the variances of the estimators, the QCR bound. For a given probe state, if $F_{Q\, i,j}=0$, $i\neq j$, the lower bound becomes
$\Sigma^2=\Tr(F^{-1}_Q)=(F_{Q\, 1,1})^{-1}+(F_{Q\, 2,2})^{-1}$. It also quantifies the joint information content about the parameters in the probe state.

In this manner, we study not only the information extraction through our measurement but also explore the information availability in the probe states themselves, leading to a thorough analysis of the problem.

In~\cite{vidrighin2014joint}, it was demonstrated that the simultaneous estimation of phase shift and diffusion using qubit quantum states, e.g., split single-photons, coherent states or N00N states, has to obey the single qubit bound (SQB). It can be expressed by a trade-off relation which holds true for such probes and separable measurements
\begin{align}
\Upsilon\leq 1
\end{align}

Moreover, for qubit probe states, the SLDs corresponding to the two parameters do not commute which implies that the QCR bound is saturated by incompatible measurements. However, in particular, the equatorial qubit states, satisfy the commutation condition (Eq.~\eqref{eqn:comm_rel})~\cite{vidrighin2014joint}. Thus, the QCR bound can be achieved asymptotically through collective measurements on multiple copies of these states. As the SQB is much tighter than the QCR limit, $\Upsilon \leq 2$ , designing a more precise estimation strategy that approaches the QCR bound more closely but does not require collective entangled measurements, requires using different probe states. 

With respect to our gHB probe states, we numerically verified that for a large class of states, the off-diagonal elements of the FI and the QFI matrices vanish.
Thus, the use of the quantifiers, namely the trade-off quantity $\Upsilon$ and the minimum absolute error $\Sigma^2$ for our analysis is justified. Additionally, for the same class of states, we numerically computed the commutator $[\mathcal{L}_{\phi},\mathcal{L}_{\Delta}]$ and the commutation condition and interestingly found that only the latter vanishes (similar to equatorial qubit states).
This implies that the collective measurements on gHB states would asymptotically saturate the QCR bound.

\subsection{Phase diffusion noise}\label{sec:phasediffusion}

\begin{figure*}[htbp]
	\centering
	\includegraphics[width=160mm]{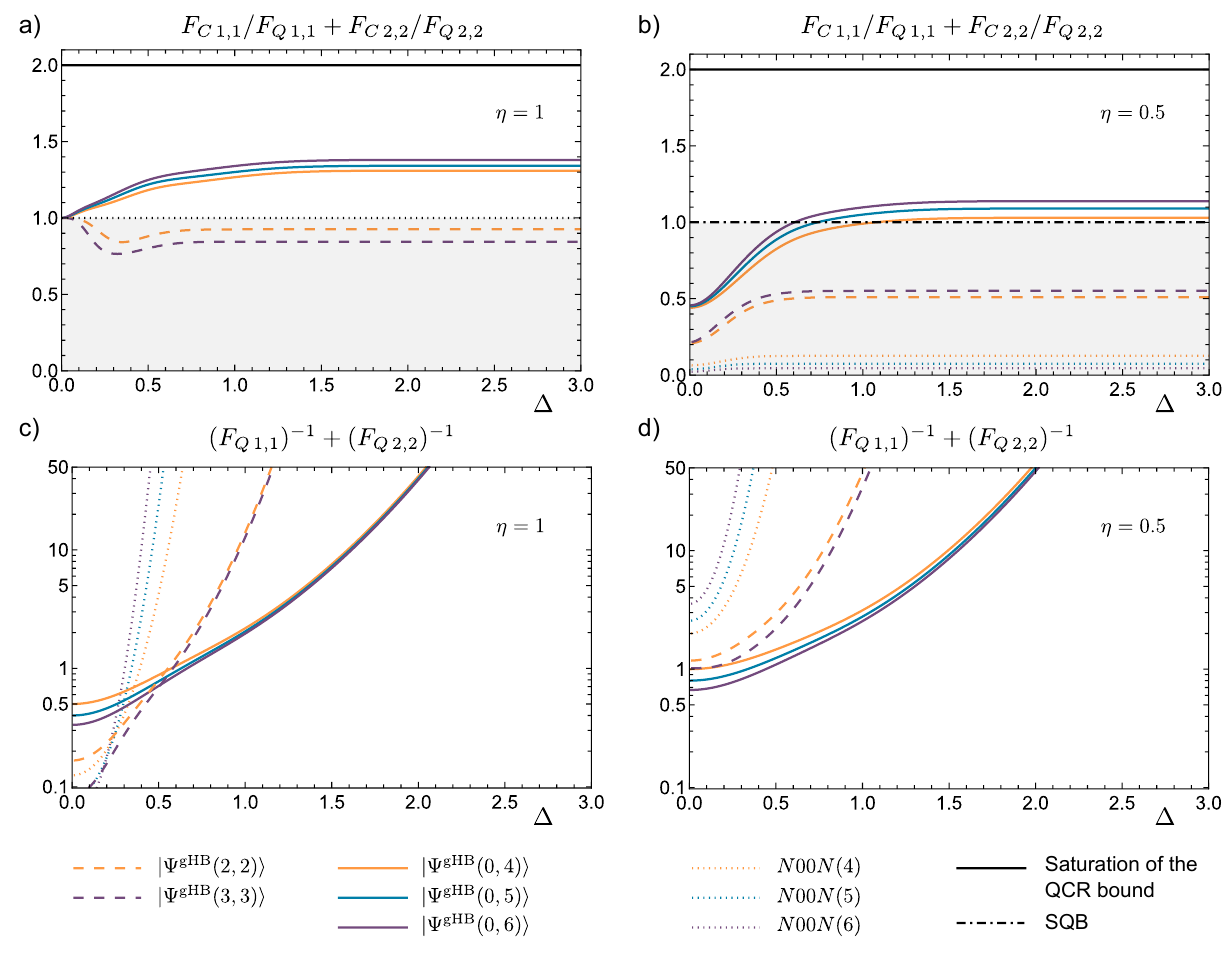}
    \caption{\small{\textit{Top row:} The trade-off quantity, considering the double homodyne measurement on gHB probe states and N00N states, is plotted as a function of the phase diffusive parameter, $\Delta$. The states with number of photons, $N=4,5,6$, are shown. Panel a) represents the lossless case and panel b) corresponds to $50\%$ photon losses. Note that the measurement extracts the highest amount of information about the parameters from the $|\Psi_{\mathrm{gHB}}(0,N)\rangle$ states across all values of $\Delta$ in both panels. \textit{Bottom row:} The QCR bound is plotted as a function of $\Delta$ for the same states considered in the top row, presented as a log plot. Panel c) illustrates the lossless case, while panel d) depicts the case with $50\%$ photon losses. Note that, in panel c), the $|\Psi_{\mathrm{gHB}}(0,N)\rangle$ states possess the highest amount of information only for $\Delta>0.6$. However, in panel d), they possess the highest amount of information for all values of $\Delta$.}}
\label{fig:wrtdelta}
\end{figure*}

We assess the performance of joint estimation of phase and phase diffusion using gHB probe states in the MZI with double homodyne measurements, by comparing it with the results obtained using HB and N00N probes in the same setup. The results are depicted in Fig.~\ref{fig:wrtdelta}. We are particularly interested in states with low total number of photons, as they are experimentally easier to prepare. Furthermore, photon losses are inevitable in any experimental setup. Hence, we incorporate them by inserting additional beam splitters of equal reflectivities at the output arms of the MZI just after the phase-shift operation. Now, the combined effects of the phase diffusive noise and the photon losses result in an output state which is highly mixed (see Section~\ref{sec:methodsB} for details).
As the FI and the QFI are functions of the output state, obtaining an analytical closed-form expression for them is challenging. This problem has been tackled using several numerical approaches \cite{demkowicz2009quantum,escher2012quantum,demkowicz2012elusive,thekkadath2020quantum,chabuda2020tensor,szczykulska2017reaching}. Additionally, for the probe states with total number of photons greater than 2, analytical computation becomes infeasible due to the rapidly increasing dimensionality of the density matrices rendering their diagonalization hard. Therefore, we resort to numerical computation of the FI and the QFI for the gHB probe states.

Firstly, we note that both the FI and the QFI matrices are independent of the phase $\phi$ (see Section~\ref{sec:methodsC} for details). Thus, the top row in Fig.~\ref{fig:wrtdelta} depicts the trade-off quantity $\Upsilon$ as a function of only the phase diffusion $\Delta$ for the lossless case in panel a), and for the case of $50 \%$ of losses in panel b). We have also indicated the SQB and the QCR bound which serve as references to additionally gauge the performance of our probe states. From panel a), we infer that carrying out the double homodyne measurement on the family of the MZI input states $|\Psi_{\mathrm{gHB}}(0,N)\rangle$ for $N=4,5,6$, violates the SQB for all values of phase diffusive noise. We have chosen specific values of $N$ such that the violation of the SQB is large enough to be observed. Furthermore, we note that violations occur even for states with $N<4$, which are relativity easier to prepare experimentally. However, it is worth noting that these states yield lower violations compared to those with $N>4$. In fact, the higher the value of $N$, the closer to the saturation of the QCR bound one can get (see Section~\ref{sec:methodsD}). We also note that while the N00N states saturate the SQB, the curves obtained for the HB states are below it (grey area). We also find that the curves for gHB and HB states attain their asymptotic values for $\Delta>1.2$.

Remarkably in panel b), we can find that even in the presence of $50 \%$ photonic losses, the performance of the double homodyne measurement on $|\Psi_{\mathrm{gHB}}(0,N)\rangle$ states is still superior to that of HB states and N00N states across all values of phase-diffusive noise. In contrast to panel a), the presence of losses significantly reduces the effectiveness of using the double homodyne measurement on N00N states in comparison to HB states. Furthermore, we have checked that for a fixed value of $N$, the $|\Psi_{\mathrm{gHB}}(0,N)\rangle$ state yields the maximum violation of the SQB among all other partitions (see Section~\ref{sec:methodsE}).

The observations resulting from panels a) and b) are specific to the choice of double homodyne detection as a measurement following the MZI. However, it is also crucial to identify the metrologically resourceful states among our probe states by assessing their information content. This will elucidate the performance of the probe states, whose information content can be fully extracted by individual measurements corresponding to the optimal estimation of $\phi$ and $\Delta$. We begin our assessment by utilizing the quantity $\Sigma^2$ which provides the minimum on the sum of the variances of the estimators.

The bottom row of Fig.~\ref{fig:wrtdelta} depicts $\Sigma^2$ for the lossless case -- panel c), and for $50 \%$ of losses -- panel d). From panel c), we conclude that the QCR bound for the $|\Psi_{\mathrm{gHB}}(0,N)\rangle$ family of states is lower than the corresponding bounds for the HB and the N00N states for diffusion values $\Delta>0.6$. As a result, the $|\Psi_{\mathrm{gHB}}(0,N)\rangle$ family of states contain higher information about the parameters than the HB and the N00N states. 

In panel d), with $50\%$ photon losses, we observe that the $|\Psi_{\mathrm{gHB}}(0,N)\rangle$ family of states retains higher information content than the HB and N00N states across all values of phase diffusive noise.

As a summary of our two-fold analysis, we infer that the joint estimation of phase and phase diffusion using the $|\Psi_{\mathrm{gHB}}(0,N)\rangle$ family of states is robust to both phase diffusion and photon losses. This aligns with the performance of the double homodyne measurement in extracting information from these family of states. 

\subsection{Number of photons in probe states}

\begin{figure*}[htbp]
    \centering
    \includegraphics[width=160mm]{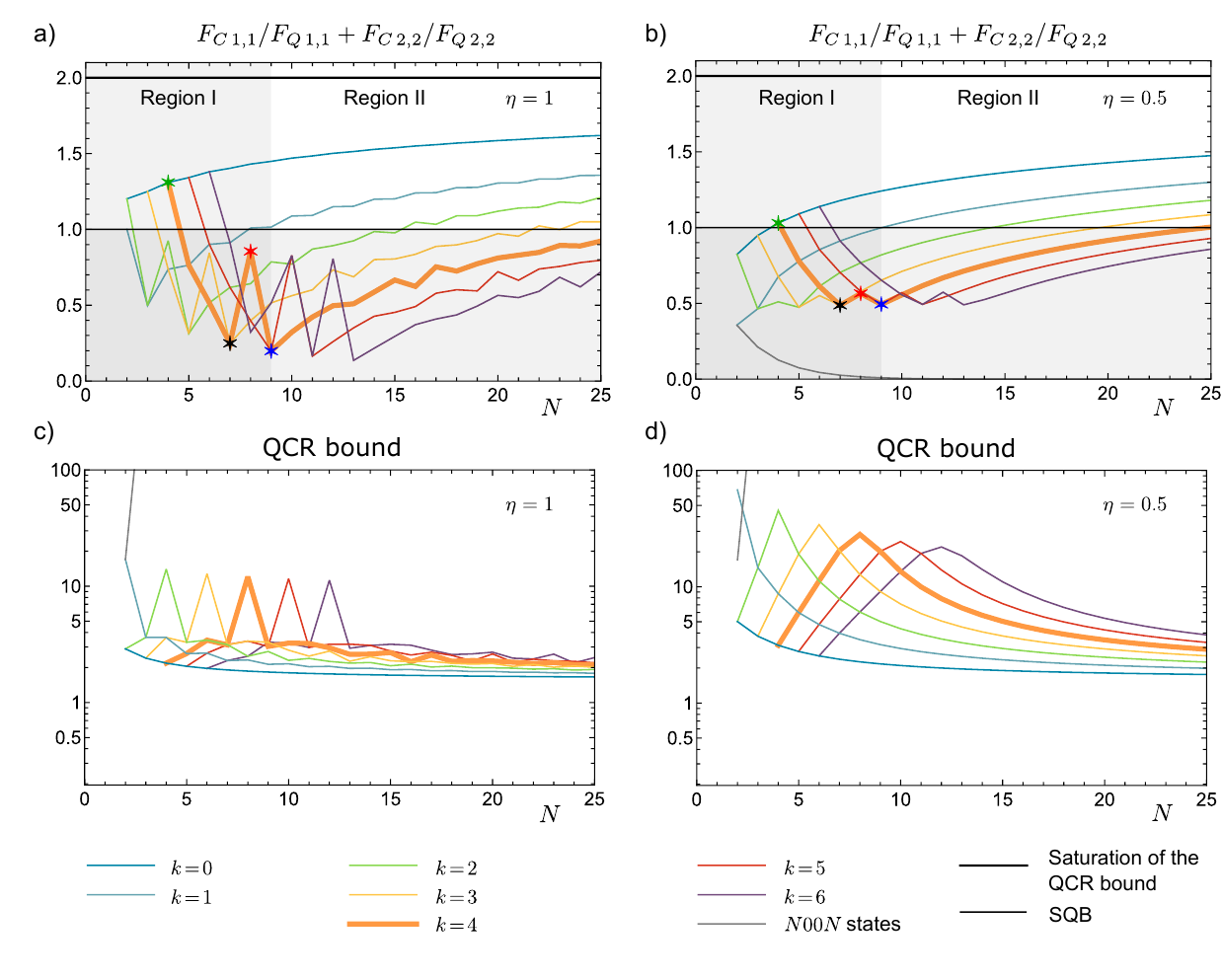}
    \caption{\textit{Top row:} The trade-off quantity, considering the double homodyne measurement on the family of gHB states $|\Psi_{\mathrm{gHB}}(k,N-k)\rangle$; $k\in[0,6]$, and N00N states, is plotted as a function of the number of photons, $N$. In the saturation region (i.e., $\Delta>1.2$ in Fig.~\ref{fig:wrtdelta}), the value $\Delta=5$ is chosen. Panel a) represents the lossless case and panel b) corresponds to $50\%$ photon losses. Note that the measurement extracts the highest amount of information from the family of gHB states with $k=0$ across all values of $N$ in both the panels. The plots can be divided into two regions of distinct characteristics, namely, Region $\mathrm{I}$ and Region $\mathrm{II}$. Considering the gHB family with $k=4$ (highlighted in orange), Region $\mathrm{I}$ is characterized by the existence of sharp points corresponding to the gHB states:  $|\Psi_{\mathrm{gHB}}(4,0)\rangle$ (green star), $|\Psi_{\mathrm{gHB}}(4,3)\rangle$ (black star), $|\Psi_{\mathrm{gHB}}(4,4)\rangle$ (red star), and $|\Psi_{\mathrm{gHB}}(4,5)\rangle$ (blue star). However, Region $\mathrm{II}$ is marked by steadily increasing curves with occasional fluctuations. \textit{Bottom row:} The QCR bound is plotted as a function of $N$ for the same states and the value of $\Delta$ considered in the top row, presented as a log plot. Note that the family of gHB states with $k=0$ possess the highest amount of information for all values of $N$ in both the panels, thereby fully aligning with the top row.}
    \label{fig:wrtphotonnum}
\end{figure*}

Even though the $|\Psi_{\mathrm{gHB}}(0,N)\rangle$ family of states achieves maximum violation of the SQB, we are yet to explore the extent of violation achieved by the other probe states, namely,  $|\Psi_{\mathrm{gHB}}(k,N-k)\rangle$ for $k\neq0$. This would provide a comprehensive analysis of the precision limits of the joint estimation through the double homodyne measurement on various probe states within the gHB family. The results are presented in Fig.~\ref{fig:wrtphotonnum}.

With this in mind, we compute the trade-off quantity in the large diffusion region where the values corresponding to all partitions of $N$ attain saturation (see Section~\ref{sec:methodsE}). In the top row of Fig.~\ref{fig:wrtphotonnum}, by setting $\Delta{=}5$, we plot the trade-off quantity with respect to $N$ for the families: $|\Psi_{\mathrm{gHB}}(k,N-k)\rangle$ with $k\in[0,6]$. We have analyzed the cases $\eta= 1$ in panel a) and $\eta=0.5$
in panel b) by marking the SQB and QCR limit as references.

In panel a), we observe that the maximum violation of the SQB is demonstrated by the states with $k=0$ among those with $k\in[0,6]$ for all values of $N$. While for other states, the violation of the SQB occurs only in certain regions of $N$. Nevertheless, a common trend among all states is that increasing the value of $N$ brings one closer to the saturation of the QCR bound or results in a greater violation of the SQB. Specifically, in region II, we infer that as we increase the value of $k$, the value of $N$ at which a particular curve intersects the SQB also increases.

From panel b), we can infer that even in the presence of $50\%$ photon losses, similar features as in the lossless case are preserved. The only exception is that the double homodyne measurement on N00N states no longer saturates the SQB and performs much worse compared to the $|\Psi_{\mathrm{gHB}}(k,N-k)\rangle$ family of states.

Similarly to the two-fold analysis in Section~\ref{sec:phasediffusion}, we next investigate the metrological resourcefulness of our probe states and N00N states. Thus, in the bottom row of Fig.~\ref{fig:wrtphotonnum}, we plot $\Sigma^2$ as a function of  $N$,  again by setting $\Delta=5$ in panel c) ($\eta=1$) and in panel d) ($\eta=0.5$).

From panel c), we can infer that the QCR bound for the $|\Psi_{\mathrm{gHB}}(k,N-k)\rangle$ family of states is lower than the corresponding bound for the N00N states across all values of $N$. Consequently, the information content of the $|\Psi_{\mathrm{gHB}}(k,N-k)\rangle$ states is higher than that of the N00N states. In particular, the maximum information content is possessed by the states with $k=0$ among those with $k\in[0,6]$ for all values of $N$.

In panel d), for the case of $50\%$ losses, we observe similar features as in panel c) further emphasizing the robustness of $|\Psi_{\mathrm{gHB}}(k,N-k)\rangle$ states to both phase diffusive noise and photon losses in comparison to the N00N states.

In summary, Fig.~\ref{fig:wrtphotonnum} underscores the impressive resilience of a broad class of gHB states, namely, the $|\Psi_{\mathrm{gHB}}(k,N-k)\rangle$ states for the joint estimation of phase and phase diffusion under highly noisy conditions. The corresponding observations made coincide with the performance of the double homodyne measurement in extracting information from these states.

\section{Discussion}\label{sec:discussion}

One of the features that characterize the trade-off curves in the top row of Fig.~\ref{fig:wrtdelta} is the existence of the saturation region for $\Delta>1.2$.
This implies that the double homodyne measurement cannot extract more information from a given probe state after a certain cut-off value of phase diffusion denoted as $\Delta_{\mathrm{cutoff}}$. Hence, its value is different for different gHB and HB states (see Section~\ref{sec:methodsE}). Incidentally, in \cite{vidrighin2014joint}, the existence of the saturation region was not reported, as their trade-off curves were plotted only up to $\Delta=1$.

Our findings support two observations from ~\cite{vidrighin2014joint} when one considers only the states with $N{=}\:6$ in the top row of Fig.~\ref{fig:wrtdelta}. Firstly, in the lossless case, the double homodyne measurement performs better with N00N states than with HB states. Secondly, in the presence of $50\%$ losses, the measurement performs better with HB states than with N00N states. Considering the gHB state $|\Psi_{\mathrm{gHB}}(0,6)\rangle$, in the lossless case, we achieve a sensitivity gain of $44.97\%$ compared to the N00N state. In the presence of $50\%$ losses, we achieve a sensitivity gain of $106.53\%$ compared to the HB state. Hence, the sensitivity obtained through the double homodyne measurement on the gHB state has more than doubled compared to the same with the HB state in the lossy case. Moreover, we know that for gHB states, $\mathrm{Tr}(\rho_{\phi,\Delta} [\mathcal{L}_{\phi}, \mathcal{L}_{\Delta}])=0$, but $[\mathcal{L}_{\phi}, \mathcal{L}_{\Delta}]\neq0$. Thus, in principle, the limit of 2 can be achieved asymptotically through collective measurements.

Again, in the top row of Fig.~\ref{fig:wrtdelta}, we can also infer that, in the absence of phase diffusion, i.e., for $\Delta=0$, the double homodyne measurement optimally extracts information about phase for all the probe states we have considered. This implies that the measurement is optimal not only for probe states that can be modeled as equatorial qubit states (as mentioned in Section~\ref{sec:setting}) but also for higher-dimensional probe states. However, in the presence of phase diffusion, the double measurement becomes an incompatible measurement.

Additional violations of the SQB have been reported in \cite{vidrighin2014joint}, employing collective measurements on two-qubit states and general projective measurements on the HB state. However, devising an experimental setup for these cases poses a substantial challenge. In contrast, our observed violations, achieved through double homodyne measurements on gHB states, act as a suitable platform for an experimental implementation.

In the top row of Figs.~\ref{fig:wrtdelta} and \ref{fig:wrtphotonnum}, the reference line $\Upsilon=2$ corresponds to the information extraction by a certain type of collective measurements. This sets the benchmark to assess the information extraction of the double homodyne measurement on various probe states. Analogously, in the bottom row of Figs.~\ref{fig:wrtdelta} and \ref{fig:wrtphotonnum}, we need a reference line to evaluate the information content of the probe states. In principle, this line is obtained by calculating the QCR bound for the optimal probe state, which contains the maximum information available for extraction about both parameters. Nevertheless, we have not included it because, in general, determining the optimal probe state in a multiparameter setting is still an open problem. It is worth noting that some key initial insights in this direction were provided in \cite{albarelli2022probe}. Therefore, as the value of $N$ increases, our probe states contain more information, approaching that of the optimal state.

Examining the top row of Fig.~\ref{fig:wrtphotonnum}, we observe that among the states with $k\in[0,6]$, those with $k=0$ require the fewest photons to surpass the SQB, specifically when $N{=}\:2$. Therefore, double homodyne measurement on the $|\Psi_{\mathrm{gHB}}(0,N)\rangle$ states are best suited for extracting higher amounts of information with a low number of photons, in a resource-efficient manner.

\section{Conclusion}\label{sec:conclusion}

We have studied the problem of multiparameter estimation of phase and phase diffusion using experimentally implementable probe states. We have approached the problem with a twofold analysis: first, by assessing how much information the double homodyne measurement extracts, and second, by evaluating the information content inherent in the probe states themselves. 

As a result, we have identified that the gHB states are a good candidate both in terms of achieving greater precision of joint estimation as well as being robust to photon losses. Furthermore, these characteristics have been demonstrated to be achieved by gHB states with low number of photons which make them even more experimentally accessible. Among the family of gHB states and the N00N states, the ones which are created by inputting all the photons in one arm of the interferometer, $|\Psi_{\mathrm{gHB}}(0,N)\rangle$, have been shown to achieve the best precision with respect to the value of the phase diffusive noise and the number of photons. 

Our work directly extends the results of \cite{vidrighin2014joint}, where HB states have been demonstrated to perform better compared to the N00N states in the presence of losses. The use of gHB states allowed us to explore the performance of states even with an unequal number of photons in the interferometer arms, thereby giving rise to a wide variety of probe states. We have illustrated a significant violation of the SQB using experimentally feasible states and a separable measurement. This contrasts with the corresponding violations demonstrated in \cite{vidrighin2014joint}, where collective measurements were applied on two-qubit states and general projective measurements were used on HB states, i.e., cases that are experimentally hard to implement. Additionally, we have also revealed the existence of the saturation region characteristic of the trade-off curves for all the gHB states. As part of our future work, we search for collective measurements on gHB states that could potentially saturate the QCR bound.

\section{Methods}

\subsection{Holevo-Cramér-Rao bound}\label{sec:methodsA}

A bound stronger than the QCR that takes into account the measurement incompatibility present in the problem we consider, is given by the Holevo-Cramér-Rao (HCR) bound. For the special case of D-invariant models to which our study belongs, HCR reads~\cite{holevo2011probabilistic,hayashi2005asymptotic} 
\begin{align}
    \Tr(\mathcal{G}\gamma)\ge \frac{1}{M}[\Tr(\mathcal{G}F_Q^{-1})+ \lVert \sqrt{\mathcal{G}}F_Q^{-1} W F_Q^{-1}\sqrt{\mathcal{G}}\rVert_{1}].
\end{align}
where $\mathcal{G}$ is a positive semi-definite cost matrix, $W_{i,j}=\Tr(\rho_{\theta}[\mathcal{L}_i,\mathcal{L}_j])$ and $\lVert \cdot \rVert_1$ is the trace norm.

\subsection{Phase diffusion and photon losses}
\label{sec:methodsB}

Photon losses are modelled using fictitious beam splitters of reflectivities $\eta_{a}$ and $\eta_{b}$ at the two arms of the MZI after the application of the phase-shift operation. If $k$ and $l$ are the number of photons lost from modes $a$ and $b$ respectively, the output gHB state after being subjected to the phase-shift operation, photon losses, and the phase diffusion operation reads
\begin{multline}
\rho_{\phi,\Delta}^{\mathrm{gHB},\mathrm{loss}}(n,N-n)={}\\
{}=\begin{aligned}[t]\sum_{p,q=0}^{N}\sum_{k=0}^{N^{'}}\sum_{l=0}^{N^{''}}
&\tilde{\mathcal{C}}_N(n,p,q)\times{}\\&{}\times\ket{p-k, N-p-l}\bra{q-k, N-q-l},\end{aligned}
\label{eqn:lossyghbstate}
\end{multline}
where $\tilde{\mathcal{C}}_N(n,p,q)=\mathcal{A}_N(n,p)\mathcal{A}_N(n,q)e^{-i(p-q)\phi - \frac{\Delta^2}{2}(p-q)^2}\break\sqrt{B^p_{kl}B^q_{kl}}$, $B^p_{kl}=\binom{p}{k}\binom{N-p}{l}\eta^{p-k}_a(1-\eta_a)^{k}\eta^{N-p-l}_b(1-\eta_b)^{l}$, $N^{'}=\min(p,q)$, and $N^{''}=\min(N-p,N-q)$. $B^p_{kl}$ quantifies the reduction of the probability amplitude due to losses. We note that Eq.~(\ref{eqn:lossyghbstate}) is a highly mixed state. Furthermore, considering general pure two-mode
input state with definite photon number $N$ in a lossy MZI, the optimal input states have been obtained numerically by exploiting the convexity property of QFI while excluding phase diffusion \cite{demkowicz2009quantum}. Unfortunately, with the inclusion of phase diffusion, the convexity property no longer holds and hence, it becomes much harder to find the optimal state. 

\subsection{The FI and the QFI matrices are independent of phase for gHB states}\label{sec:methodsC}

Here, we provide a set of arguments for the phase independence of the FI and the QFI matrices. We note that $ \rho^{\mathrm{gHB}}_{\phi,\Delta}(n,N-n)$ (Eq.~\ref{eqn:outputghbstate}) can be rewritten in the new basis $\ket{p,N-p}_{\phi}=e^{-ip\phi}\ket{p,N-p}$ by absorbing the phase factor into the Fock basis. Thus,  
\begin{multline}    
  \rho^{\mathrm{gHB}}_{\phi,\Delta}(n,N-n)={}\\{}=\sum_{p,q=0}^{N}\mathcal{C}'_{N}(n,p,q)\ket{p,N-p}_{\phi}\bra{q,N-q}_{\phi},
\label{eqn:newbasis}
\end{multline}
where $\mathcal{C}'_{N}(n,p,q)=\mathcal{A}_N(n,p)\mathcal{A}_N(n,q)e^{- \frac{\Delta^2}{2}(p-q)^2}$. 

The double homodyne measurement is characterized by the joint measurement of quadratures $\hat{X}$ and $\hat{P}$ with outcomes $x$ and $p$ respectively. Thus, the associated joint probability distribution reads $p_{\phi,\Delta}(x,p)=\Tr(\rho^{\mathrm{gHB}}_{\phi,\Delta}\Pi(x,p))$, where $\Pi(x,p)=\ket{\nu(x,p)}\bra{\nu(x,p)}$ is the double homodyne POVM such that $\int dx dp \,\Pi(x,p)=\mathds{1}$ and $\Pi(x,p)\geq0$. Using Eq.~(\ref{eqn:newbasis}) and also writing $\Pi(x,p)$ in the new basis, it can be easily shown that $p_{\phi,\Delta}(x,p)$ is independent of $\phi$. Hence, it follows that the FI matrix elements $F_{C \, \phi,\Delta}=\int dx dp \frac{1}{p_{\phi,\Delta}}\frac{\partial}{\partial \phi} p_{\phi,\Delta} \frac{\partial}{\partial \Delta} p_{\phi,\Delta}$ are also independent of $\phi$. For further details about the exact form of $\Pi(x,p)$, one may refer to the supplemental material of \cite{vidrighin2014joint}. 

Writing $\rho^{\mathrm{gHB}}_{\phi,\Delta}$ in the eigenbasis, i.e., $\rho^{\mathrm{gHB}}_{\phi,\Delta}=\sum_{p}\lambda_p\ket{p,N-p}_{\phi}\bra{p,N-p}_{\phi}$, where $\{\lambda_p\}$ are the eigenvalues, the SLDs are defined as\break $\mathcal{L}_{\phi\,(\Delta)} \!=\! \sum_{p,q}\frac{2\bra{p,N-p}_{\phi}\partial_{\phi\,(\Delta)}\ket{q,N-q}_{\phi}}{\lambda_p+\lambda_q}\, \ket{p,N-p}_{\phi} \bra{q,N-q}_{\phi}$; $\partial_{\phi\,(\Delta)}=\partial/\partial \phi\,(\Delta)$. Again, using Eq.~(\ref{eqn:newbasis}) and the SLDs, it is straightforward to show that the QFI matrix elements, $F_{Q \, \phi,\Delta} = \frac{1}{2}\mathrm{Tr}(\rho^{\mathrm{gHB}}_{\phi,\Delta} \{\mathcal{L}_\phi, \mathcal{L}_\Delta\})$ are independent of $\phi$.

\subsection{Violation of the SQB by the family of states, $|\Psi_{\mathrm{gHB}}(0,N)\rangle$}\label{sec:methodsD}

\begin{figure*}[htbp]
	\includegraphics[width=80mm]{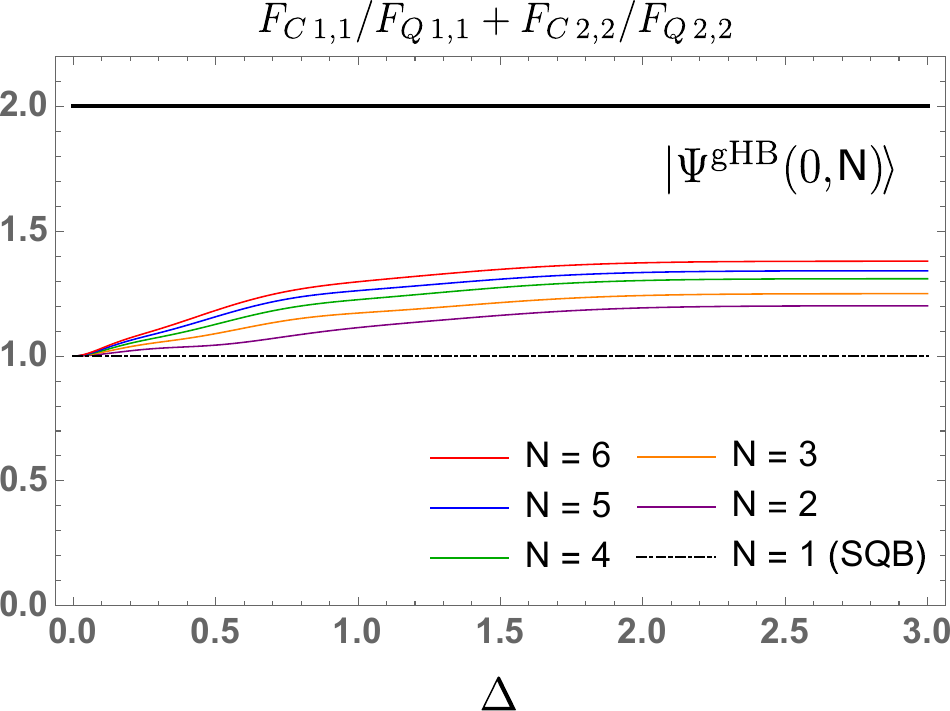}
    \caption{\small{The trade-off quantity is plotted as function of $\Delta$ for the family of gHB states, $|\Psi_{\mathrm{gHB}}(0,N)\rangle$; $N\in[1,6]$. The SQB and the saturation value of the QCR bound are taken as references. Note that the higher the value of $N$, the greater the violation of the SQB.}}
\label{fig:tradeoff0N}
\end{figure*}

In this section, for $\delta=N$, we analyze the behaviour of the trade-off quantity for various values of $N$ with respect to the phase diffusion parameter. Again, the lossless case is considered in Fig.~\ref{fig:tradeoff0N} and we can find that greater violations of the SQB occur at higher values of $N$. Although states with $N>6$ are of theoretical interest from the perspective of sensitivity gain, they are highly susceptible to experimental imperfections. It is important to note that violations occur even for states with $N<4$.


\onecolumngrid
    
\twocolumngrid


\subsection{Maximum violation of the SQB and saturation values of the trade-off quantity}\label{sec:methodsE}

\begin{figure*}[htbp]
    \centering
    \includegraphics[width=160mm]{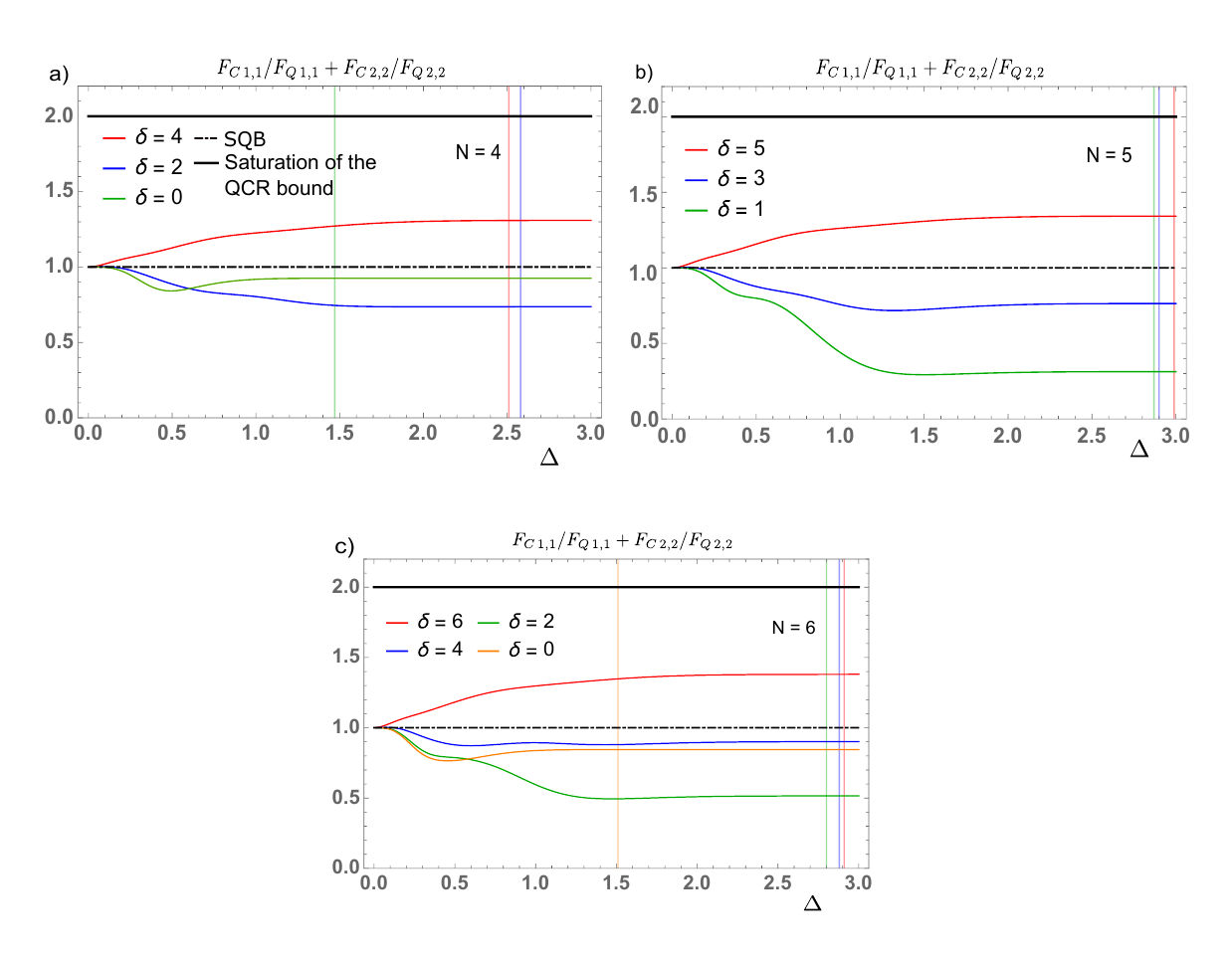}
    \caption{\small{The trade-off quantity is plotted as a function of $\Delta$. In each panel, for a given value of $\delta$, a vertical grid line represents the corresponding cutoff value, $\Delta_{\mathrm{cutoff}}$. Note that the maximum violation of the SQB is demonstrated by states with $\delta=N$.}}
    \label{fig:tradeoffpartitions}
\end{figure*}

In this section, for a fixed value of $N$, we analyze the behaviour of the trade-off quantity for various partitions of $N$ with respect to the phase diffusion parameter. The partitions are denoted as $\delta=N-2n$. The lossless case is considered in Fig.~\ref{fig:tradeoffpartitions}, and we can clearly observe that the maximum violation of the SQB is demonstrated by the gHB states with $\delta=N$, i.e., the state $|\Psi_{\mathrm{gHB}}(0,N)\rangle$, among all other partitions. Particularly, $\delta=0$ corresponds to the HB state. Also, we find that the saturation region for the trade-off quantity exists for all values of $\delta$. However, as stated earlier in Section~\ref{sec:discussion}, the cut-off value, $\Delta_{\mathrm{cutoff}}$, after which the saturation occurs is different for each state. A suitable value of $\Delta_{\mathrm{cutoff}}$ must be chosen in this region such that all the gHB states regardless of the values of $N$ and $\delta$ attain saturation. This is the basis for choosing $\Delta_{\mathrm{cutoff}}=5$ in Fig.~\ref{fig:wrtphotonnum}.
\newpage


\onecolumngrid
    
\twocolumngrid


\begin{acknowledgments}
J.J. and M.S. were supported by the National Science Centre ``Sonata Bis'' project No. 2019/34/E/ST2/00273. M.S. was supported by the European Union’s Horizon 2020 research and innovation programme under the Marie Skłodowska-Curie project ``AppQInfo'' No. 956071, and the QuantERA II Programme that has received funding from the European Union’s Horizon 2020 research and innovation programme under Grant Agreement No 101017733, project ``PhoMemtor'' No. 2021/03/Y/ST2/00177. M.B. acknowledges support from the Italian Ministry of University through the project PRIN22-RISQUE-2022T25TR3 and Dipartimento di Eccellenza 2023-2027. J.J. thanks Juan Camilo L\`opez Carr\~eno for fruitful discussions. 
\end{acknowledgments}

\vfill


\bibliography{refs}


\end{document}